\documentclass[12pt,a4paper]{article}
\pdfoutput=1
\usepackage{jheppub}
\usepackage[titletoc,title]{appendix}
\usepackage{bm}
\usepackage{amsmath}
\usepackage{amssymb}
\usepackage{ifsym}
\usepackage{amsthm}
\usepackage{amsfonts}
\usepackage{ccaption}
\usepackage{booktabs}
\usepackage{mathrsfs}
\usepackage{braket}
\usepackage{subfig}
\makeatletter
\@addtoreset{equation}{section}

\makeatletter
\renewcommand\section{\@startsection {section}{1}{\z@}%
                                   {-3.5ex \@plus -1ex \@minus -.2ex}%nn
                                   {2.3ex \@plus.2ex}%
                                   {\normalfont\large\bfseries}}
\renewcommand\subsection{\@startsection{subsection}{2}{\z@}%
                                     {-3.25ex\@plus -1ex \@minus -.2ex}%
                                     {1.5ex \@plus .2ex}%
                                     {\normalfont\bfseries}}

  \captionnamefont{\bfseries}
  \captiontitlefont{\small\sffamily}
  \captiondelim{: }
  \hangcaption
  \renewcommand{\figurename}{Fig.}

\definecolor{point}{rgb}{0.7,0.,0.3}

\newcommand{\be}{\begin{equation}}
\newcommand{\ee}{\end{equation}}
\newcommand{\ben}{\begin{eqnarray}\displaystyle}
\newcommand{\een}{\end{eqnarray}}

\newcommand{\p}{\partial}

\newcommand{\trho}{\ensuremath{\tilde{\rho}}}
\newcommand{\tz}{\ensuremath{\tilde{z}}}

\newcommand{\calc}{\mbox{${\cal C}$}}

\newcommand{\calo}{\mbox{${\cal O}$}}

\title{Cusp Anomalous dimension and rotating open strings in AdS/CFT}

\author{R. Esp\'indola and J. Antonio Garc\'ia}

\affiliation{Departamento de F\'isica de Altas Energ\'ias, Instituto de Ciencias Nucleares\\
Universidad Nacional Aut\'onoma de M\'exico,\\
Apartado Postal 70-543, Ciudad de M\'exico, 04510, M\'exico}
\emailAdd{ricardo.espindola@correo.nucleares.unam.mx, garcia@nucleares.unam.mx}

\abstract{In the context of AdS/CFT we provide analytical support for the proposed duality between a Wilson loop with a cusp, the cusp anomalous dimension, and the meson model constructed from a rotating open string with high angular momentum. This duality was previously studied using numerical tools in \cite{Caron-Huot-Henn}. Our result implies that the minimum of the profile function of the minimal area surface dual to the Wilson loop, is related to the inverse of the bulk penetration of the dual string that hangs from the quark--anti-quark pair (meson) in the gauge theory.}

\begin{document}
%\begin{flushright} \small{DCPT-15/01} \end{flushright}

\maketitle

\flushbottom
\renewcommand{\thefootnote}{\arabic{footnote}}
% ______________________________________
\newpage
\section{Introduction and result}
By a clever combination of modern methods of scattering amplitudes \cite{scatering}, the old Regge  scattering amplitudes approach (Regge trajectory) \cite{Gribov} and taking advantage of the dual conformal momentum space symmetry of the planar ${\cal N}=4$ SYM theory in four dimensions, the authors of \cite{Caron-Huot-Henn} (see also \cite{Co-He-Ma-Se}) obtained a duality between  the higher angular momentum state for a given energy $E$ in terms of the cusp anomalous dimension $\Gamma_{cusp}$ and  $\theta$ the cusp angle of the Wilson line with a cusp given by \cite{Co-He-Ma-Se, hnss}, 
\ben
J+1=-\Gamma_{cusp}(\theta),\qquad E^2=4m^2\cos^2 \theta/2. \label{conf-duality}
\een
where $J$ is the angular momentum, $m$ is the mass and $E$ the energy.
This duality was scrutinized in the perturbative regime
as well as in the strong coupling limit using AdS/CFT, reproducing results of the meson spectrum \cite{mateos} using numerical methods. Unfortunately, the authors in \cite{Caron-Huot-Henn}  were unable to find an analytic relation between the string model of the meson, constructed from an open rotating string as the dual of the meson (massive quark anti-quark pair) and the cusp anomalous dimension using the Wilson loop approach. The aim of this paper is to fill this gap and give an explicit analytic expression
 that relates the parameters of these  two very different approaches to the problem of finding the meson spectrum. We will work in the context of AdS/CFT (semiclassical approximation) so the window that we will explore is the strong coupling regime in the field theory side.

The central result of our paper is that we can calculate the energy and the angular momentum of the meson model in {\em closed} form in terms of only one parameter $\tz_0$ that correspond to the maximum penetration of the dual string profile in AdS. The result is
\ben
E(\tz_0)/m_q=2\cos (\theta(\tz_0)/2),\label{energy-z0-I}
\een
\ben
J(\tz_0)+1=\frac{\sqrt\lambda}{4\tz_0\sqrt{1+\tz_0^2}}~ \null_2F_1(\frac12,\frac32,2,-\frac{1}{{1+\tz_0^2}}),\label{ang-mo-z0-I}
\een
where
\ben
\theta(\tz_0)=2K\int_0^\infty \frac{d\xi}{(1+\xi^2)\sqrt{(1+\tz_0^2+\xi^2)(2+\tz_0^2+\xi^2)}},\label{theta-z0-I}
\een
is the cusp angle, $E$ is the energy of the quark, anti-quark pair, $m_q$ the quark mass, $J$ the angular momentum and $K=\tz_0\sqrt{1+\tz_0^2}$. $~_2F_1$ is the hypergeometric function.

Using these formulas we can reconstruct the information encoded in the meson spectrum. In particular,  the energy and angular momentum in parametric form with parameter $\tz_0$. It is usual to represent this result in a parametric $E-J$ plot. The plot was reported in \cite{mateos} and reproduced in \cite{Caron-Huot-Henn}. In the context of AdS/CFT duality we provide here a version of this plot (see Fig.~4). In this figure we show that in fact the numerical integration of the meson model is equivalent to our results. Fig.~4 shows the Regge trajectories for different values of $\lambda$, the `t Hooft coupling, obtained by numerical integration of the equations of motion for the string profile as compared with  the analytic result given by our relations (\ref{energy-z0-I}) and (\ref{ang-mo-z0-I}) in terms of the parameter $\tz_0$ that measures the maximum penetration of the string profile in the bulk. Our results  reproduce the well-known asymptotic analytical information as well as predict new information for the next-to-next to leading order for $E$ and $J$ and in fact for the complete functions $E$ and $J$. 
The information provided by our formulas is equivalent to solve a differential equation for the profile of the string that as far as we know was not solved in analytic form yet. In fact,  using the Wilson loop with a cusp angle $\theta$ and then relating $\theta$ to the energy and the angular momentum with $\Gamma_{cusp}$ the cusp anomalous dimension,  through the duality (\ref{conf-duality}) we can recover all the information encoded in the string profile as well as the energy and angular momentum from the bulk of the meson model. 
Recently another very interesting observable, namely the length of the string in the meson model, was related to the anomalous dimension of twist two operators in the field theory side \cite{Iancu-Braga}.

We will review the construction of the central concepts relevant to our work and then present support for our main results (\ref{energy-z0-I}) and (\ref{ang-mo-z0-I}). 
A crucial point relies on the identification of a parameter of the string profile in the meson model to a corresponding parameter in the Wilson loop calculation.
 The identification of this parameter will be given by observing that the corresponding analytical expressions that come from the duality (\ref{conf-duality}), 
through $\Gamma_{cusp}$ and the cusp angle $\theta$ associated with the angular momentum and the energy respectively, match the corresponding expressions of the meson model through the duality (\ref{conf-duality}). To that end the central arguments behind the duality (\ref{conf-duality}) will be reviewed. The cusp anomalous dimension in the context of AdS/CFT can be calculated from a cusped Wilson loop but also can be related to Regge trajectory from scattering amplitudes in ${\cal N}=4$ SYM (see Sect. 2). This is an instance of the scattering amplitudes/ Wilson loop correspondence. 

In section 3 we will review the basic ingredients of the meson model constructed from a rotating string in AdS and extract analytic information through the analysis of interesting asymptotic limits. These analytical results are crucial to develop our relation between a Wilson loop with a cusp and meson model. We will present them in a form that is well suited for our general proposes. 

In section 4 we present a compact presentation of the dual conformal symmetry of planar ${\cal N}=4$ SYM amplitudes and their use to relate the cusp anomalous dimension and the cusp angle to the angular momentum and the energy respectively. We refer the reader to the original references for details.

In sections 5 and 6, we present our results and a dictionary that gives analytical support to the duality (\ref{conf-duality}) in the context of AdS/CFT. Finally in section 7, we present some final comments  and provide some ideas for future work.

  \section{Cusp anomalous dimension}

\subsection{$\Gamma_{cusp}$ from Wilson loop}

We will recall the construction of the cusp anomalous dimension from the definition of the Wilson loop in the context of AdS/CFT. The notation and conventions are given in \cite{Dorn} and \cite{Gross}. For details about the calculation of the cusp anomalous dimension we refer the reader to the original articles.

In the context of the AdS/CFT, the expectation value
of a Wilson loop in the gauge theory is given by the action of a string bounded by the curve at the 
boundary of space \cite{Maldacena-Wilson},
\be
\langle W[C]\rangle
=\int_{\partial X={\cal C}} \mathscr{D} X\, \exp( -\sqrt{\lambda} S[X]),
\label{defWL}
\ee
where $S[X]$ is the string action. For large $\lambda$,
we can estimate the path integral by the steepest descent method.
Consequently the expectation value of the Wilson loop
is related to the area $A$ of the minimal surface
bounded by ${\cal C}$ as
\be
\langle W \rangle \approx \exp ( - \frac{\sqrt{\lambda}}{2 \pi} A(\mathcal{C}) ) .
\label{WL-AdS}
\ee
The construction starts by considering
Euclidean AdS space in Poincar{\' e} coordinates
\be\label{metric-cusp}
ds^2=\frac{1}{z^2}(dx^\mu dx_\mu +dz^2)
\ee
where the conformal boundary is at $z=0$.

The computation of the Wilson loop in $AdS$ requires an IR
regularization, since
the area of the minimal surface ending at the boundary
of $AdS$ is infinite due to the factor $z^{-2}$ in the metric.
Thus, in order to make sense of the ansatz (\ref{WL-AdS}),
we need to regularize the area.  The standard procedure to regularize the area  $A_{\epsilon}$ is based on cutting off that part of the surface at which $z<\epsilon$. 
On the gauge theory side, the Wilson loop requires
regularization in the ultraviolet.
According to the UV/IR relation, the IR cutoff $\epsilon$ in AdS should be identified with the UV cutoff in the gauge theory.
If $\ell$ is the length of the Wilson loop contour, the regularized area $A_\epsilon$ is defined by
$$A_\epsilon=\frac{\ell}{\epsilon}+A_{ren}+O(\epsilon).$$
In the presence of a cusp a new logarithmic divergence appears. 
$$A_\epsilon=\frac{\ell}{\epsilon}+\Gamma_{cusp}(\theta)\log\epsilon+A_{ren}+O(\epsilon).$$

Consider a loop formed by two half-lines with cusp angle $\theta$ on the two dimensional plane $(x_1,x_2)$ defined by the metric (\ref{metric-cusp}). 
In this case we can obtain an analytic expression for the minimal surface area whose boundary is the corresponding loop. We use polar coordinates to parametrize the worldsheet,
\be
x_1 = \rho \ \cos \varphi, \  \ \ x_2 = \rho \ \sin \varphi.
\ee 
Using the conformal symmetry $z\to\lambda z$, $x_\mu\to\lambda x_\mu$, we can assume that the minimal surface is described by the anzats \cite{Gross} 
\be
z(\rho,\varphi)=\frac{\rho}{f(\varphi)},
\label{ansatz}
\ee
with boundary conditions $z(\rho,0) = z(\rho, \theta)=0$.  This symmetry under dilatation is crucial for our proposes because it is a guide to relate the parameters of the Wilson loop with the corresponding parameters in the meson model. We need to relate an invariant under dilatations in the two models. A natural candidate is the function $f(\varphi)$.

With this factorization of the $\rho$  and $\varphi$ dependence, the minimal surface condition, i.e. the equation of motion, becomes an ordinary differential equation for $f(\varphi)$ with boundary conditions $f(0)= f(\theta)=\infty$. 
In what follows we will not need the explicit form of this differential equation. The integration can be formally induced if we notice that the translation symmetry ($\varphi \mapsto \varphi - \alpha$, $\alpha$ an infinitesimal constant angle) implies the conserved quantity,
\be
\mathcal{H} = \mathcal{L} - f' \frac{\p \mathcal{L}}{\p f'} = \frac{f^4 + f^2}{\sqrt{f^4 + f^2 + (f')^2}}.
\ee
 From the symmetry of the profile $f$ has a minimum at $\varphi = \theta/2$ where $\theta$ is the cusp angle. Using this condition and the above conserved quantity we can write
 \be
 K=f_0 \sqrt{1+f_0^2}, \ \ \ \ \ f_0 = f(\theta / 2). 
 \label{Evalue}
 \ee
Hence, the integration yields
\be
\theta= 2K \int\limits_{f_0}^\infty \frac{df}{\sqrt{(f^4+f^2)^2 - K^2(f^4+f^2)}},
\label{thetaint}
\ee
which fixes the relation between $f_0$ and the cusp angle $\theta$. Indeed, $\theta(f_0)$ is a monotonically decreasing function of $f_0$, $\theta(0)=\pi$ and $\theta(\infty) = 0$ (see Fig. 6 blue line).

The regularized area defined with cutoffs $z=\rho/f(\varphi)>\epsilon$ and $\rho< L$ is
$$A_{\epsilon,L}=\int d\rho \ d\varphi \ \frac{\sqrt{f^4 + f^2 + (f')^2}}{\rho}$$
$$=\frac{2L}{\epsilon}+\Gamma_{cusp}\log\frac{\epsilon}{L}+A_0(\theta)+\ldots$$
where the dots denote terms vanishing for $\epsilon\to 0$ and
$$\Gamma_{cusp}=2f_0-2 \int_{f_0}^\infty\left(\sqrt{\frac{f^4+f^2}{f^4+f^2-K^2}}-1\right)df.$$
$A_0$ is a regular term that we will not need in what follows.

The substitution $f^2=f_0^2+\eta^2$ gives \cite{Kruczenski}
$$\Gamma_{cusp}(\theta)= \int\limits_{-\infty}^{\infty} d\eta \Big( 1 -  \sqrt{\frac{1+\eta^2+f_0^2}{1+\eta^2+2 f_0^2}} \Big).$$
The first term  inside the integral is an infinite subtraction constant which makes the area finite. We can thus identify the logarithmic divergence coefficient with the cusp anomaly. A closed form of this integral can be obtained for $\Gamma_{\rm cusp}$, in terms of a hypergeometric function \cite{Dorn},
\be
\Gamma_{\rm cusp}(\theta) = \frac{\pi}{2} \frac{f_0^2}{\sqrt{1+f_0^2}}~~ _2F_1\Big (\frac{1}{2},\frac{3}{2},2,\frac{-f_0^2}{1+f_0^2}\Big )~.
\label{cusp}
\ee

 This relation between the cusp anomalous dimension and the parameter $f_0$ and the corresponding relation between $\theta$ and $f_0$ given by
 \be
 \theta(f_0) = 2 K \int\limits_0^\infty \dfrac{d\eta}{(\eta^2 + f_0^2)\sqrt{(\eta^2+f_0^2+1)(\eta^2+2f_0^2+1)}}, 
\label{thetaintK}
\ee
where the change of variable $ f^2 = f_0^2+\eta^2$ was applied, will be very important in what follows.

\subsection{$\Gamma_{cusp}$ from scattering amplidudes}

$\Gamma_{cusp}(\theta)$ can also be calculated from scattering amplitudes of massive particles in ${\cal N}=4$ SYM. It can be extracted as the coefficient in front of the IR divergence of the scattering amplitudes. Thanks to a powerful dual conformal symmetry it is possible to show that the scattering amplitudes depends only on two invariant ratios ${\cal M}(u,v)$. In the limit $u<<1$ the coefficient of the IR divergence of the amplitude can be identified with the coefficient of the UV divergence of the Wilson loop with a cusp \cite{Co-He-Ma-Se}
$$\ln {\cal M}(u,v)\to (\ln u) \Gamma_{cusp}(\lambda, \theta)+\ldots$$
From Regge theory, in the same limit the amplitud scales as \cite{hnss}
$${\cal M}(u,v) \sim  u^{-(j(s)+1)}$$
where $j(s)$ is the leading Regge trajectory. 
These different physical interpretations of the same amplitude ${\cal M}(u,v)$ relates $\Gamma_{cusp}$ to the Regge trajectory $j(s)$\footnote{ Here we write the duality in the AdS/CFT context, that is the reason why the factor $\sqrt{\lambda}$ appears in front of $\Gamma_{cusp}$. The duality can also be used in the perturbative regime and the dependence on $\lambda$ is quite different.}
\be
j(s)+1 = - \left( - \frac{\sqrt{\lambda}}{2 \pi}\right)  \Gamma_{\it cusp}(\theta), \ \ \ \ s=4 m^2 \cos^2 \left(  \frac{\theta}{2} \right) 
\label{duality}
\ee
where $s=E^2$ is the energy in the CM frame, $j(s)$ is a Regge trajectory and  $\Gamma_{\it cusp}$ is the cusp anomalous dimension associated with a Wilson loop with cusp angle $\theta \in \left[0, \pi \right] $. 
This relation has been checked up to three loops in Ref. \cite{Co-He-Ma-Se}, to which we refer the reader for more details. 

In particular (\ref{duality}) can be used to study the massive quark anti-quark potential. This perspective open the way to compare the results that come from (\ref{duality}) with other models where the massive quark anti quark potential is involved. One of this models is the meson model \cite{mateos} that we will review in the next section. From this meson model we can extract non perturbative information about the spectra of the meson $J(E)$ and then compare this information with $\Gamma_{cusp}$ calculated as in subsection 2.1 using a Wilson loop. It turn out the this two very different perspectives provide numerical support to the duality (\ref{duality}) \cite{Caron-Huot-Henn}.

The aim of our work is to provide analytical support to the validity of this relation (\ref{duality}) in the context of the AdS/CFT correspondence. In section 3, we will recover the asymptotic information of the meson model in two asymptotic analytical limits and in section 4 we will give complete support for the validity of this duality for the complete interval, not just the asymptotic limits considered in sections 3.1 and 3.2.

\section{Meson spectrum from a rotating string}

The  AdS/CFT  duality is nowadays a basic tool to study some aspects of strongly coupled gauge theories by using a semiclassical gravity solution in asymptotically $AdS$ space-time \cite{malda}. The best studied case is the duality between ${\cal N}=4$ SYM theory in four dimensions at strong coupling and a weakly coupled gravitational theory in five dimensional $AdS$. 
It is well known that to include fields in the {\em fundamental} representation of SU(N) (quarks) in AdS/CFT, one must introduce probe D7-branes in $AdS$ \cite{Karch-Katz}. The dual gauge theory is the ${\cal N} = 2$ SYM theory with massive quarks. 
The beta function of this theory is $\beta \propto \lambda^2N_f/N$, where $N_f$ is the number of flavour D7 branes and $N$ is the number colour D3 branes \cite{kirsch}. This beta function tends to zero for $N_f$ small, fixed  't Hooft coupling $\lambda$ and $N \rightarrow \infty$ (planar limit), such that the theory remains conformal in this limit.

Mesons of low spin are described as fluctuations of the D7-branes. High spin mesons are represented by semiclassical rotating open strings attached to the D7-branes \cite{mateos} (see also \cite{Erdmenger} for a review).

The presence of the D7-branes introduce a mass scale in the duality, the quark mass $m_q$, corresponding to the position of the branes along the radial direction of AdS5. The mass $m_q$ of the quark in the gauge theory is related to the separation distance $L=R^2/z_{D7}$ between D3-branes and D7-branes by the relation
\be
\label{mq}
m_q = \frac{R^2}{2\pi \alpha' z_{D7}}
\ee
where $z_{D7}$ is the position of the D7 branes in the bulk\footnote{In this work we will ignore the backreaction of the D7 branes \cite{kirsch}. An interesting question not considered here is if this backreaction can still be adapted to fit in the framework of the duality (\ref{conf-duality}).}
The introduction of a mass scale breaks the invariance under dilatations. Nevertheless we have a new isometry that implies
that the observables are appropriate rescaled variables. For example the energy is not an invariant but $E/m_q$  indeed is an invariant. The separation between the string endpoints on the D7-brane is not an invariant but we can find a rescaled variable that plays the role of the meson size in the gauge theory.

The string configuration, can be constructed as a solution of the Nambu Goto action with appropriate boundary conditions starting from the metric 
\ben
ds^2 = \frac{R^2}{z^2} \left(
-dt^2 + d\rho^2 + \rho^2 d\theta^2 + dz^2 \right)\  ,\label{metric-meson}
\een
where the spatial coordinates are $(\rho,\theta)$. As stated in \cite{mateos} a high-spin meson can be represented as a semiclassical rotating string with a profile $\rho = \rho(z), \theta = \omega t$.  The maximum value of $\rho$ occurs  at $z = z_0$. We are interested in U shaped solutions where the end points of the string hangs from de D7 brane located at $z = z_{D7}$. As remarked above this parameter is a measure of the quark mass $m_q$. A very massive meson implies that the D7-brane is near the boundary. 

As discussed in \cite{mateos}, it is convenient to use the dimensionless coordinates $\tilde \rho \equiv \omega \rho$ and $\tilde z\equiv\omega z$. 

Choosing the static gauge $\tau=t$ and $\sigma=\tilde z$, the Nambu-Goto action takes the form
\ben
S = - \frac{R^2 \omega}{\pi \alpha'} \int d t \int^{\tilde z_0}_{\tilde z_{D7}}d \tilde z\,
\frac{1}{\tz^2} \sqrt{\left(1- \trho^2 \right)
\left( \trho' \,^2 + 1 \right)}\ ,\label{NGS}
\een 
Here a prime denotes a derivative with respect to $\tz$. The string angular momentum is 
\ben
J=\frac{\partial {L}}{\partial\omega}=\frac{R^2}{\pi\alpha'}\int^{\tilde z_0}_{\tilde z_{D7}}d \tilde z\,
\frac{1}{\tz^2} \sqrt{\frac{ \trho' \,^2 + 1}{1- \trho^2 }}\ \label{JS}
\label{Spin}
\een 
and the meson energy 
\ben
E=\omega\frac{\partial L}{\partial \omega} - L =\frac{R^2\omega}{\pi\alpha'}\int^{\tilde z_0}_{\tilde z_{D7}}d \tilde z\,
\frac{1}{\tz^2} \sqrt{\frac{ \trho' \,^2 + 1}{1- \trho^2 }}\ .\label{ES}
\label{Energy}
\een 
The string profile $\tilde\rho(\tilde z)$ is determined by solving the corresponding equations of motion with the boundary condition 
$$\frac{d\tilde \rho}{d\tilde z}\Big|_{\tilde z=\tilde z_{D7}}=0$$
i.e. the string ends orthogonally on the D7-brane.

The conditions at $\tilde z_0$, are
$$\tilde \rho(\tilde z_0)=0, \qquad \frac{d\tilde \rho}{d\tilde z}\Big|_{\tilde z=\tilde z_0}\to-\infty.$$
The equation of motion and these boundary conditions determine the maximal string penetration $\tilde z_0$ as function of the position $\tilde z_{D7}$ of the D7-brane, and the string profile $\tilde \rho(\tilde z)$ for $\tilde z_{D7}\leq\tilde z\leq\tilde z_0$. 

Unfortunately, the equations of motion with these boundary conditions cannot be solved by analytical methods and the only recourse at hand is to approximate the profile $\tilde \rho(\tilde z)$ using numerical methods. 

The calculation can be drastically simplified by observing the following relation between string parameters
 $\tilde z_{D7}$ and $\tilde z_0$. We found a quite simple closed form 
\ben
\tilde z_{D7}(\tilde z_0)=\frac{\calc \tilde z_0^3}{1+\calc \tilde z_0^2},\label{zd7z0pade}
\een
where $\calc = \frac{\sqrt{2}\pi^{3/2}}{\Gamma(1/4)^2}$. 
The constant ${\cal C}$ is determined by the asymptotic behavior of $\tilde z_{D7}$. This is a central result that will have an important role in what follows.

A plot of this relation is represented in Fig. 1, where the green curve is (\ref{zd7z0pade}) and we compare it to the numerical approximation (dotted line) using shooting techniques. The agreement is excellent in all the interval. 
\begin{figure}[h]
\begin{center}
\includegraphics[height=10cm,width=14cm]{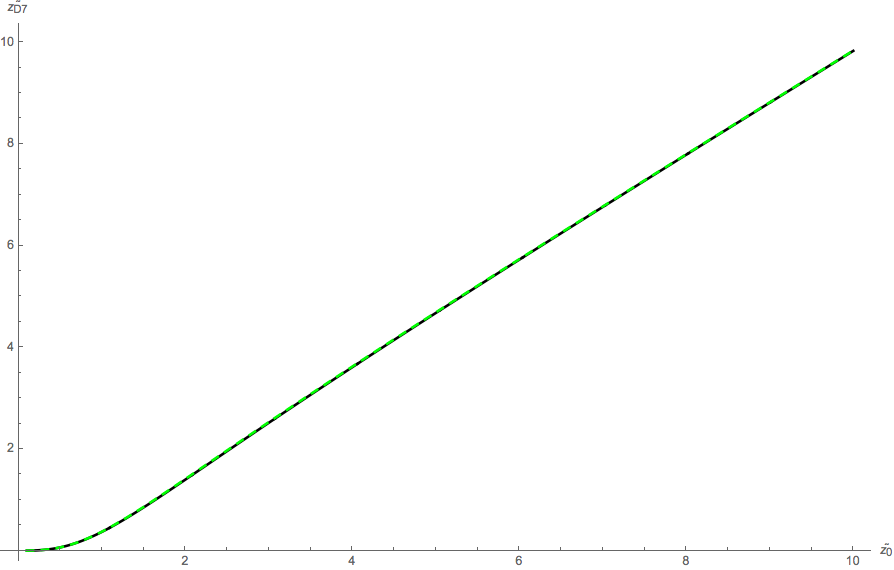}
\caption{\small\sffamily The relation (\ref{zd7z0pade}) (green) and the numerical results for $\tz_{\rm D7}(\tz_0)$ (dotted line).}
\end{center}
\end{figure}
This relation is valid through all the interval between the end points of the solution for the profile and determines the spectrum $E(\tilde z_0)$ and $J(\tilde z_0)$ as functions of only one parameter. In  the following we will choose $\tilde z_0$ for reasons that will become clear later. 
Analytic results can be obtained in the two regimes $\tilde z_0>> 1$ and $\tilde z_0<<1$. 

\subsection{$\tilde z_0<< 1$ limit}

We see from (\ref{zd7z0pade}) that in this regime 
\ben\label{zD7z0<1}
\tilde z_{D7}(\tilde z_0)\sim \calc \tilde z_0^3
\een
so $\tilde z_0\sim \omega^{1/3}$ and we conclude that the string is rotating with a very slow angular velocity $\omega\to 0$. In this circumstance we expect that analytic information can be extracted from perturbations of the static profile of the string. In fact from \cite{mateos} we have
\begin{equation}\label{EMz0<1}
E \simeq 2 m_q\left[1-\frac{\calc^2
\tz_0^2}{2}+\calo(\tz_0^4)\right]\ , \ \ \ \ \ \ \calc = \frac{\sqrt{2}\pi^{3/2}}{\Gamma(1/4)^2}
\end{equation}
From equation (\ref{Spin}), the result for the spin $J$ at leading order in $\tz_0$ is
\be\label{JMz0<1}
J\simeq \frac{R^2\calc}{\pi \alpha
\tz_0}+\calo(\tz_0).
\ee
As $\tz_0<<1$ we have $J/\sqrt{\lambda}>>1$ as expected.   
By eliminating $\tz_0$ and restoring $m_q$, one gets
\be
E = 2 m_{\rm q} - E_b \,, \ \ \ \ \ \ \  E_b = m_q \frac{\kappa^4}{4 J^2} \ \ \ \ \ \
\kappa^4 = \frac{16 \calc^4 g_s N}{\pi} \,.
\label{E-JMz0<1}
\ee
where $E_b$ is the binding energy. 
The asymptotic relations (\ref{EMz0<1}) and (\ref{JMz0<1}) are the main results of this section and will be useful in what follows. We will review now the complementary limit $\tz_0>>1$.

\subsection{$\tz_0>>1$ limit }

In the limit $\tz_0>>1$,  the relation (\ref{zd7z0pade}) take the form 
\ben
\tilde z_{D7}(\tilde z_0)\sim  \tilde z_0\label{zD7z0>1}
\een
the string profile remains very near to the position of the D7-brane.
As $\tilde z_0\sim \omega$  we conclude that the string is rotating with a high angular velocity $\omega\to \infty$.
One can check by numerical analysis that  $\tz(\rho) > \omega z_{\rm D7}\simeq\tz_0$.
The solution for the profile that we are interested in, is the near flat space solution. The string profile remains very near the D7 brane so the string is not probing the AdS spacetime.
 In this case the equations (\ref{ES}) and (\ref{JS}) give
the energy and the spin to leading order 
in $1/\omega z_{0}$ \cite{mateos}:
\ben\label{EMJMz0>1}
E \simeq \frac{\pi m_q}{ \tz_{0}}, \ \ \
J \simeq \frac{\sqrt\lambda}{4  \tz_{0}^2}\ .
\een
 Since $\tz_0 \gg 1$ we have $J \ll\sqrt{\lambda}$.
Eliminating $\tz_0$ we find
\ben\label{E-Jz0>1}
E \simeq 
\frac{\sqrt{2}\pi^{3/4} m_{\rm q}}{(g_s N)^{1/4}} \sqrt{J}\ .
\een
Therefore for $J\ll \sqrt{\lambda}$ the meson masses follow a Regge behaviour.

%***************************************

%The relation between the radial coordinate of the brane and the quark mass si given by $m_q = L/2\pi \alpha'=R^2/2\pi \alpha'  z_{\rm D7}$. Thus, fixing $z_{\rm D7}$ amounts to fixing the unique scale $m_q$ in the gauge theory. 

%Consider the high-spin meson as represented by a rotating string in the $\rho \theta$-plane so we set $\theta =\omega \tau$ where $\omega$ is the angular velocity of the string. Working in the static gauge then the Nambu-Goto action takes the form,

%\ben
%S = - \frac{R^2 \omega}{2\pi \alpha} \int d \tau d \sigma\,
%\frac{1}{\tz^2} \sqrt{\left(1- \trho^2 \right)
%\left( \trho' \,^2 + \tz' \,^2 \right)}\ .
%\een 
%where we have introduced new dimensionless time-independent string profile variables
%\ben
%\widetilde{\rho} = \omega \rho,  \ \ \ \ \  \widetilde{z}=\omega z,
%\een
%and the condition that the string endpoints are attached to the D7-brane at $z_{\rm D7}$ translates into the Dirichlet boundary condition $\tz|_{\p \Sigma}= \omega z_{\rm D7}$.
%The energy and the spin are respectively,
%\ben
%E &=& \omega\frac{\partial L}{\partial \omega} - L =
%\frac{R^2 \omega}{2\pi \alpha} \int d \sigma\,
%\frac{1}{\tz^2} \sqrt{\frac{\trho' \,^2 + \tz' \,^2}{1- \trho^2}}\ ,
%\label{Energy} \\
%J &=& \frac{\partial L}{\partial \omega} =
%\frac{R^2}{2\pi \alpha} \int d \sigma\,
%\frac{\trho^2}{\tz^2} \sqrt{\frac{\trho' \,^2 + \tz' \,^2}{1- \trho^2}}\ .
%\label{Spin}
%\een
 
 %************************

\section{Asymptotic matching between meson model and cusp anomalous dimension }

In this section we will present the main result of our paper. 
Our first question is if we can match the asymptotic results that come from the meson model, eqs. (\ref{EMz0<1}), (\ref{JMz0<1}) and (\ref{EMJMz0>1}) with the corresponding asymptotic results for the angular momentum and energy that come from the duality (\ref{duality}). The numerical evidence presented in \cite{Caron-Huot-Henn} shows that, in fact, the matching is possible, at least numerically. 
This numerical matching is very surprising because the two approaches are quite different. 
The aim of this section is to show explicitly that the match between the asymptotic results can be obtained in {\em analytical form}. The first observation is that the cusp angle and the cusp anomalous dimension depend parametrically on one parameter $f_0$ of the minimal worldsheet  dual to the Wilson loop. From the other hand the meson angular momentum and energy depend on the worldsheet parameter of the string profile dual to the meson, namely $\tilde z_0$. Notice that these two parameters are invariant under dilatations that are isometries of their respective metrics. The question can be reformulated as if we can construct a relation between these two different parameters. The answer is the affirmative and in fact, the relation is quite simple
\ben
f_0 \to \frac{1}{\tilde z_0}\label{dict}
\een

The central result of our paper is that we can calculate the energy and the angular momentum of the meson model in {\em closed} form in terms of only one parameter $\tz_0$ that is the maximum penetration of the string profile in AdS of the meson model. To obtain the result that we announced previously in the Introduction, we make use of the parameter $\tilde z_0$ in  equations (\ref{cusp}) and (\ref{thetaintK}) in place of $f_0$ and apply the duality (\ref{duality}). We quote again the result, 
\ben
E(\tz_0)/m_q=2\cos (\theta(\tz_0)/2)\label{R1}
\een
\ben
J(\tz_0)+1=\frac{\sqrt\lambda}{4\tz_0\sqrt{1+\tz_0^2}}~ \null_2F_1(\frac12,\frac32,2,-\frac{1}{{1+\tz_0^2}})\label{R2}
\een
where
\ben
\theta(\tz_0)=2K\int_0^\infty \frac{d\xi}{(1+\xi^2)\sqrt{(1+\tz_0^2+\xi^2)(2+\tz_0^2+\xi^2)}}\label{R3}
\een
and $K=\tz_0\sqrt{1+\tz_0^2}$. $~_2F_1$ is the hypergeometric function.
By using these formulas the reconstruction of the meson spectra is straightforward. In particular,  we  can write the energy and angular momentum in parametric form with parameter $\tz_0$ {\em to any order} in powers of $\tilde z_0$. A parametric plot of these basic observables is the $E-J$ graph reported in \cite{mateos} and reproduced elsewhere. Our graph (see Fig. 2) obtained with the aid of equations (\ref{R1}) and (\ref{R2}) coincides with previously reported plot using numerical methods \cite{Caron-Huot-Henn} so we provide analytical support for the proposed conformal duality (\ref{duality}) in the context of AdS/CFT.

As a check we can recover the asymptotic analytical information given in \cite{mateos} and reproduce  the asymptotic limits. Taking $\tz_0>>1$, (equivalent to take $f_0\ll 1$) we get
\be
\theta(\tz_0 ) = -  \pi   \frac{1}{\tz_0} +\pi+ \cdots.
\label{thetserie}
\ee
On the other hand, the corresponding asymptotic limit for the cusp anomalous dimension for the parameter 
 $\tz_0>> 1$ is
\be
J(\tz_0)+1 = \frac{\sqrt{\lambda}}{2\pi}(\frac{\pi}{2}\frac{1}{\tz_0^2} - \frac{7}{16} \pi \frac{1}{\tz_0^4} + \dots )  . 
\label{gammasmallf0}
\ee
 The $\tz_0 >>1$ limit corresponds to $f_0\to 0$ and then to a  cusp angle $\theta$ near $\pi$ (where no scattering take place). To leading order in $1/\tz_0$ we have
\be
\frac{E}{m} = 2 \cos \left(  \frac{\theta}{2} \right)  \approx \pi - \theta
\label{Ethetasmall} 
\ee
and
\be
\frac{E}{m} \simeq \pi \frac{1}{\tz_0}, \ \ \ \  J+1 \simeq \frac{\sqrt{\lambda}}{4} \frac{1}{\tz_0^2},
\label{Esmallf0}
\ee
by using our results (\ref{R1}, \ref{R2}) and the expansion (\ref{thetaserie}) just to leading order in the parameter $\tz_0$. Perfect agreement with the previously reported formulas is evident. We could predict the next-to-next to leading order behaviour of the meson as is clear from our closed relations.

The other interesting limit is $\tz_0<<1$. In this limit $f_0\to\infty$ and the corresponding cusp angle  $\theta\to 0$. From the expression (\ref{thetaserielarge}) we have to leading order in $\tz_0$ (see appendix A), 
 \be
 \theta(\tz_0) = \frac{2 \pi^{1/2} \Gamma\left( \frac{3}{4} \right) }{\Gamma\left( \frac{1}{4} \right) } {\tz_0},
 \label{thetaserielarge}
 \ee
For the energy, as the cusp angle is small, we can use a simple expansion 
\be
\frac{E}{m} = 2 \cos \left(  \frac{\theta}{2} \right)  \approx 2- \frac{\theta^2}{4}
  =  2 - \pi \left( \dfrac{\Gamma\left(\frac{3}{4} \right) }{\Gamma\left(\frac{1}{4} \right) }\right) ^2 {\tz_0^2} 
  = 2 -  \pi \left( \frac{2 \pi^2}{ \Gamma \left( \frac{1}{4} \right)^4}\right) {\tz_0^2} 
\label{Edualitysmallz0}
\ee
where we have used the identity
$ \Gamma \left( \frac{3}{4} \right){ \Gamma \left( \frac{1}{4} \right)} = {\sqrt{2} \pi}.$
At first sight this is not the result obtained from the analysis of the corresponding asymptotic limit from the meson  (\ref{EMz0<1}), unless
$${\cal C}^2=\pi \left( \frac{2 \pi^2}{ \Gamma \left( \frac{1}{4} \right)^4}\right).$$
Using the definition $\calc = \frac{\sqrt{2}\pi^{3/2}}{\Gamma(1/4)^2}$,
we can verify that this relation is in fact valid. So we conclude that our proposal recovers the correct analytical information that we can extract from the meson model.  For the angular momentum we obtain
\be
J(\tz_0)+1 \simeq \sqrt{\frac{\lambda}{\pi}} \frac{\Gamma\left( \frac{3}{4} \right) }{4 \Gamma\left( \frac{5}{4} \right) } \frac{1}{\tz_0} =\frac{\sqrt{2 \pi \lambda}}{\Gamma \left(\frac{1}{4} \right)^2 } \frac{1}{\tz_0},
\label{Jdualitysmallz0}
\ee
where we have used the identity $\frac{\Gamma \left( \frac{3}{4} \right) {\Gamma \left( \frac{1}{4} \right)^2}}{\Gamma \left( \frac{5}{4} \right)} = {4 \sqrt{2} \pi}$. This result coincides with the meson model using the definition of $\calc$. We stress that our relation gives much more than the asymptotic behaviour of the meson model.

To grasp the meaning behind the identification (\ref{dict}) recall that for a given cusp angle $\theta$, $f_0$ is the minimum of the function $f$ defined in (\ref{ansatz}), and by symmetry considerations it is defined by $f_0=f(\theta/2)$. Consider a plane with fixed $z$, then $f_0$ is proportional to the distance from the cusp to the minimum of the curve $\rho\sim f(\varphi)$ (see Fig. 2 and Fig. 3). $\theta$ is a monotonically decreasing function of $f_0$ starting from $\theta=\pi$, $f_0=0$ and going to  $\theta\to 0$ as $f_0\to\infty$ (see Fig. 6 blue curve). Then according to our dictionary $\tz_0\to\infty$ as $\theta\to\pi$ and $\tz_0\to 0$ as $\theta\to 0$. In the first case the behaviour  of the meson is governed by a Regge trajectory $E\sim\sqrt{J}$ and in the second case the energy is of the order $E\sim 2m_q$, the binding energy. So the cusp angle has a very nice interpretation in terms of our dictionary. When the Wilson loop is nearly smooth the behaviour of the meson model is the Regge limit and when the cusp is very spiky the meson has a nearly vanishing binding energy.

\begin{figure}[h]
\begin{center}\label{cusp1}
\includegraphics[height=10cm,width=14cm]{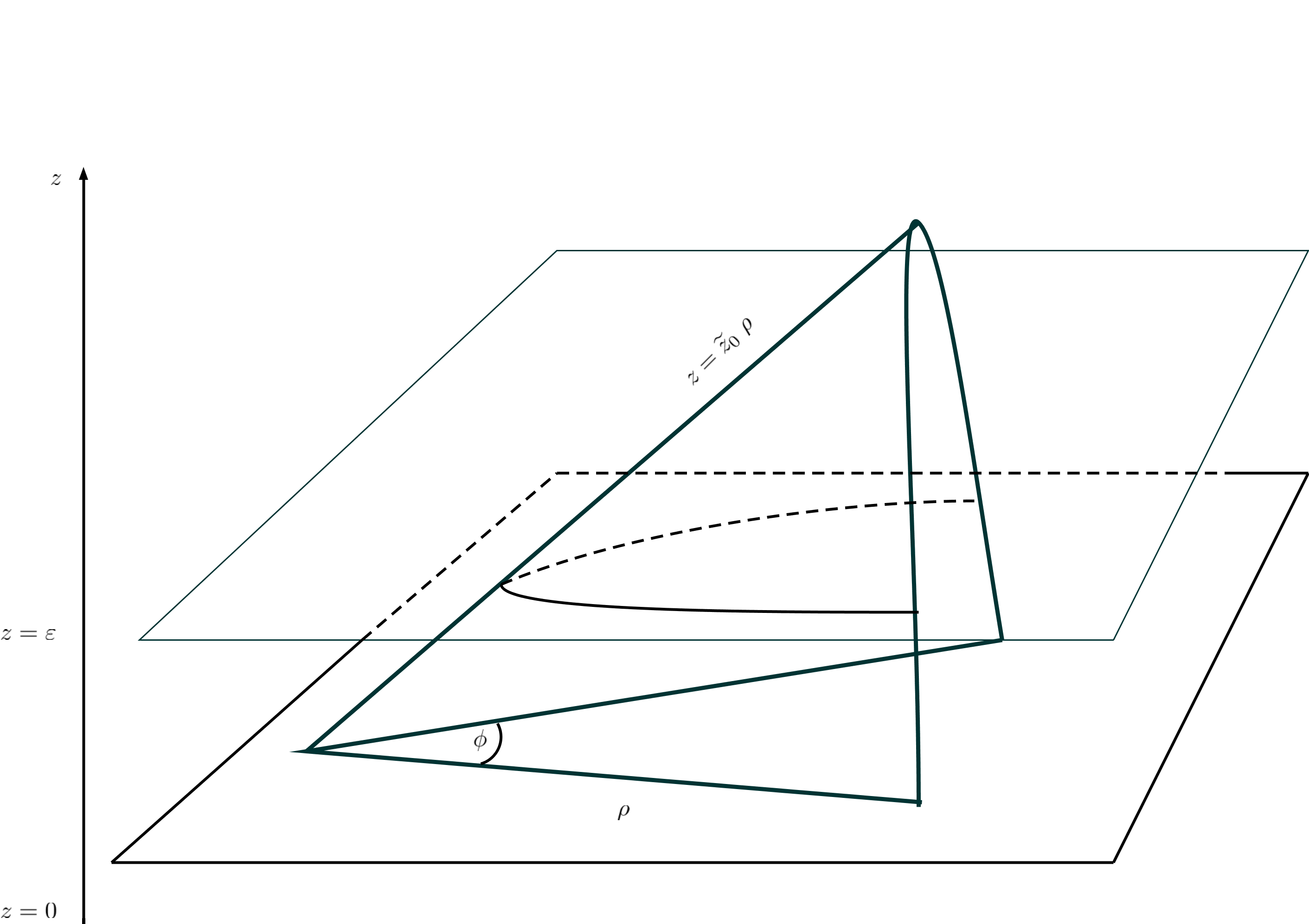}
\caption{\small\sffamily The minimal surface based on the Wlison loop with a cusp. We also show the regularizarion parameter $\epsilon$.}
\end{center}
\end{figure}
\begin{figure}[h]
\begin{center}\label{zplanecusp2}
\includegraphics[height=5cm,width=14cm]{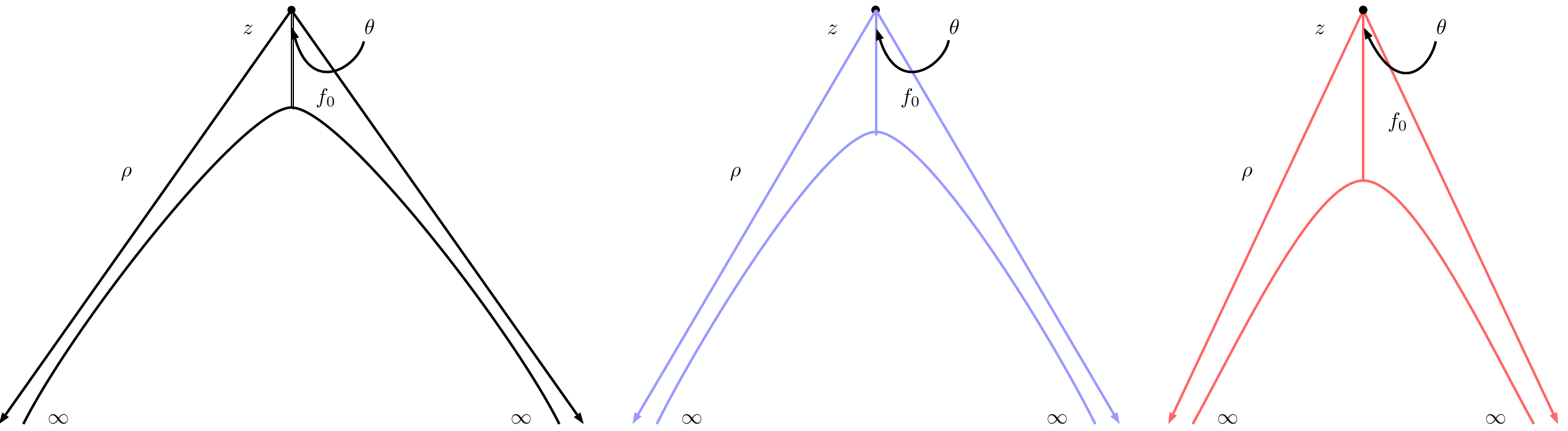}
\caption{ \small\sffamily A transverse cut of the minimal surface at $z=const.$. The geometric meaning of the parameter $f_0$ is proportional to the distance of the cusp vertex to the minimal value of the function $f(\varphi)$. We show diverse profiles of the function $f$ as we vary the cusp angle $\theta$. }
\end{center}
\end{figure}

It is also interesting to notice that in the limit $\tz_0\to\infty$ as $\theta\to\pi$ the Wilson loop is very smooth and the Bremsstrahlung function $B(N,\lambda)$ \cite{breem} can be used as an exact calculation for any $N_c$ and $\lambda$ for the cusp anomalous dimension $\Gamma_{cusp}$. The relation is
$$\Gamma_{cusp}(\theta-\pi) = B (\theta-\pi)^2 + O(\theta^4). $$

\begin{figure}[h]
\begin{center}\label{E-J}
\includegraphics[height=8cm,width=14cm]{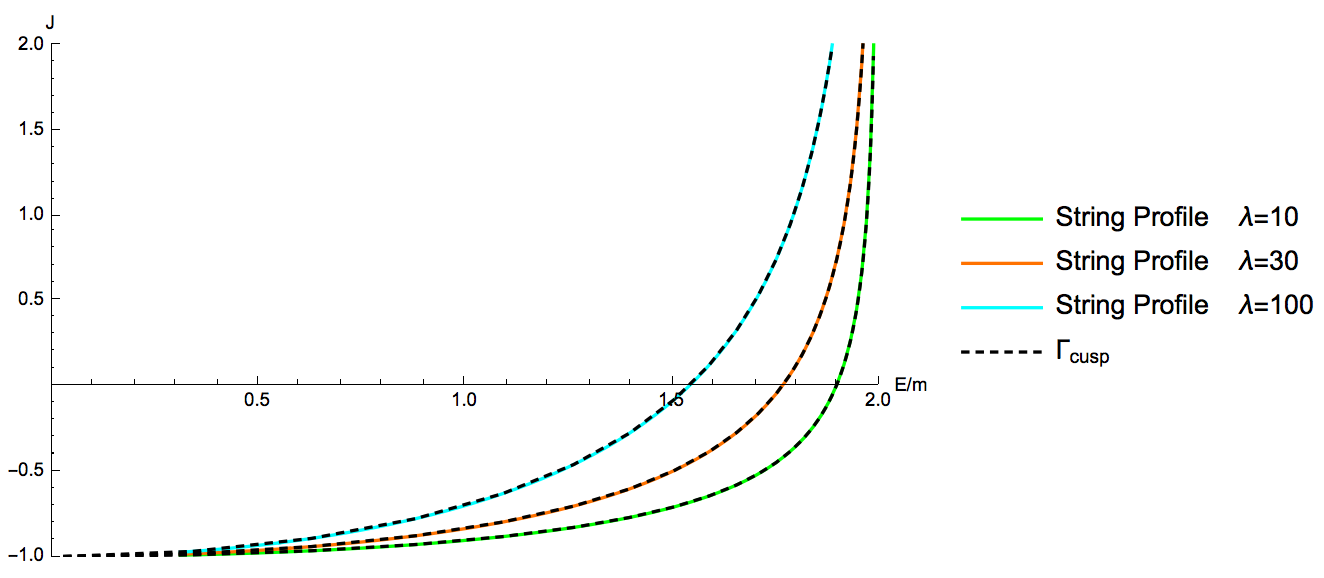}
\caption{\small\sffamily Regge trajectories of ${\cal N}=4$  SYM for $\lambda=10,30,100$. The color lines are numerical results from the meson model and the black curve is the analytical result (\ref{R1}, \ref{R2}).}
\end{center}
\end{figure}
We obtained these formulas using the results of the Wilson loop with a cusp angle $\theta$ and the conformal duality reported in section 4 by a simple application of the dictionary (\ref{dict}). The results given are valid in the context of AdS/CFT duality with planar ${\cal N}=4$ SYM theory.

%From \cite{mateos} the angular momentum dependence of the meson energy to leading order in the static string profile  $\trho(\tz)$, in the $\omega \rightarrow 0$, has the form
%
%\ben
%& \frac{E}{m}  =  2  - \dfrac{\kappa^4}{4 (J+1)^2}.\\
%& = 2 - \dfrac{4 \pi^4 \lambda}{\Gamma(\frac{1}{4})^8 (J+1)^2} \\
%& {\it where,} \ \ \ \ \ \kappa^4 = \frac{4  \lambda}{\pi^2}\calc^4 \ \ \ \& \ \ \ \ \calc = \frac{\sqrt{2}\pi^{3/2}}{\Gamma(1/4)^2}
%\label{ElargeJJ}
%\een

\section{Exploring the relation between $f_0$ and ${\tilde z_0}$ beyond the asymptotic limits}

In this section we will show that the relation between the two very different approaches, the Wilson loop with a cusp and the meson model given by the identification of the maximum string penetration ${\tilde z_0}$ and the minumun  value of the function $f(\varphi)$, $\varphi \in (0,\theta)$, $f_0=f(\frac{\theta}{2})$ with $\theta$ the cusp angle given by (\ref{dict}) is also valid for the entire interval between the asymptotic limits considered in the previous sections. 

From the fact that the asymptotic limits (\ref{Esmallf0}, \ref{Edualitysmallz0}, \ref{Jdualitysmallz0}) can be described with the same reparametrization we could argue that the reparametrization is universal, i.e. is valid for the entire interval.
But a more compelling approach is to show it with the aid of the plots in Fig.~5. In the plots (a) and (b) we show the numerical integration of the NG equation of motion for the string profile and then we integrated  numerically the equations (\ref{Spin}) and (\ref{Energy}) to obtain  $E({\tilde z_0}), J({\tilde z_0})$ for the meson model. The Fig.~5 (a) and (b) also show the corresponding asymptotic limits as reported in the original work 
\cite{mateos} and that we reproduce here in eqs.~(\ref{EMz0<1},\ref{JMz0<1}) and (\ref{EMJMz0>1}). The dotted lines in these graphs are the numerically evaluated corresponding  observable in the complete interval. 

Next, in Fig.~5 (c) and (d), we plot the corresponding graphs for the same observables $E/m_q$ and $J/\sqrt{\lambda}$ but now using the results from the Wilson loop with a cusp, the cusp anomalous dimension $\Gamma(f_0)$, and $\theta(f_0)$ and the duality (\ref{duality}). In this way we obtain the energy and angular momentum as functions of $f_0$, the minimum of the profile function $f(\varphi)$. {\em Now, the dots come from our previous graphs, evaluated at $1/f_0$ as predicted by our dictionary}. The full lines in colour comes from the evaluation of the integral for $\theta(f_0)$ and the hypergeometric function for $\Gamma_{cusp}$ given in (\ref{thetaint}) and (\ref{cusp}) respectively, and then using the duality relation to obtain  $E(f_0), J(f_0)$.

The agreement is eloquent! So we conclude that our dictionary gives a real complete description of the meson model using information from the cusp anomalous dimension and the angle of the corresponding cusp $\theta\,$\footnote{This result could come as a surprise at least by the following reasons. This is an example that in some cases the asymptotic behaviour determines the complete function for an observable in the gauge theory. In the context of AdS/CFT our experience shows that using two point Pad\'e approximants to interpolate between the asymptotic behaviour of an observable give a surprising good agreement with the exact result (when the exact result is known) and/or with the general consideration of the behaviour of the corresponding  function between the asymptotic limits under consideration \cite{Banks, unpub}.}. The case that we have at hand was scrutinized by us from using interpolation and give excellent results. Nevertheless these interpolation methods can not be used  to prove conclusively that the interpolated observables are in fact unique (we have a landscape of possible Pad\'e approximants). And/or that the behaviour of the function must be exactly controlled by the interpolation.
 
 \begin{figure}[ht]
\centering
{\subfloat[]{
\includegraphics[height=5cm,width=7cm]{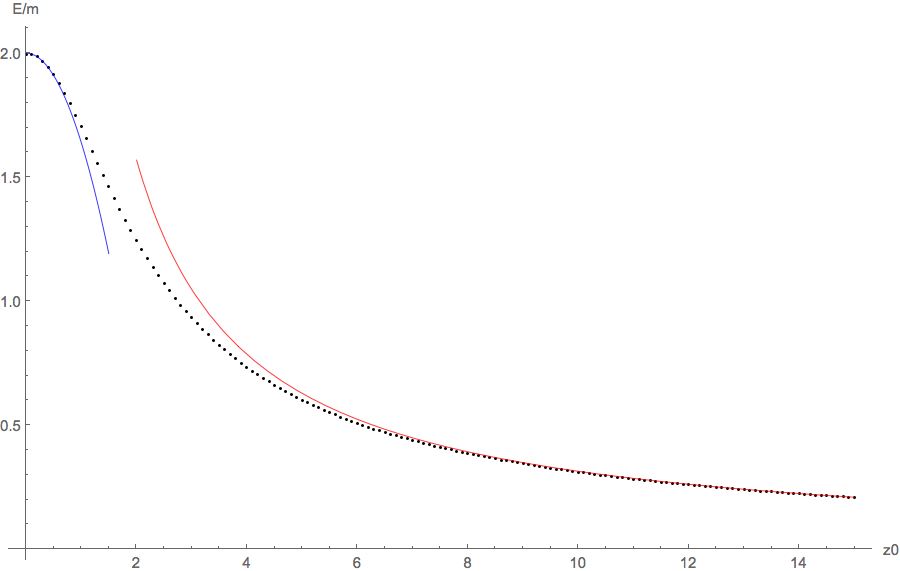}}
\quad
\subfloat[]{
\includegraphics[height=5cm,width=7cm]{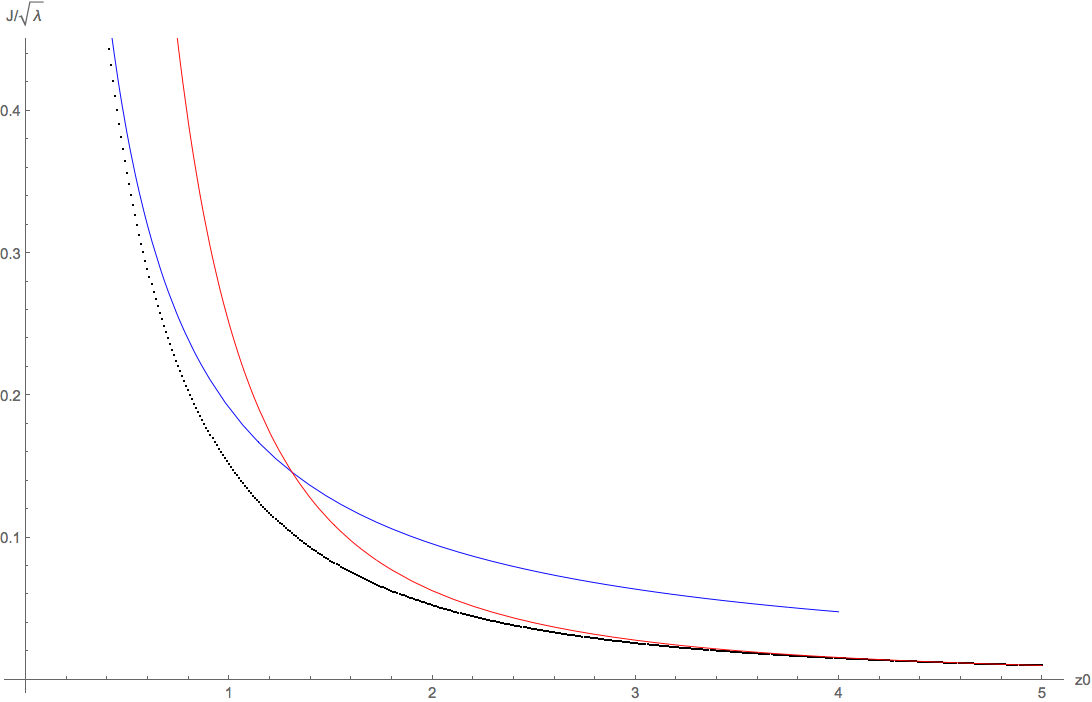}}}
{\subfloat[]{
\includegraphics[height=5cm,width=7cm]{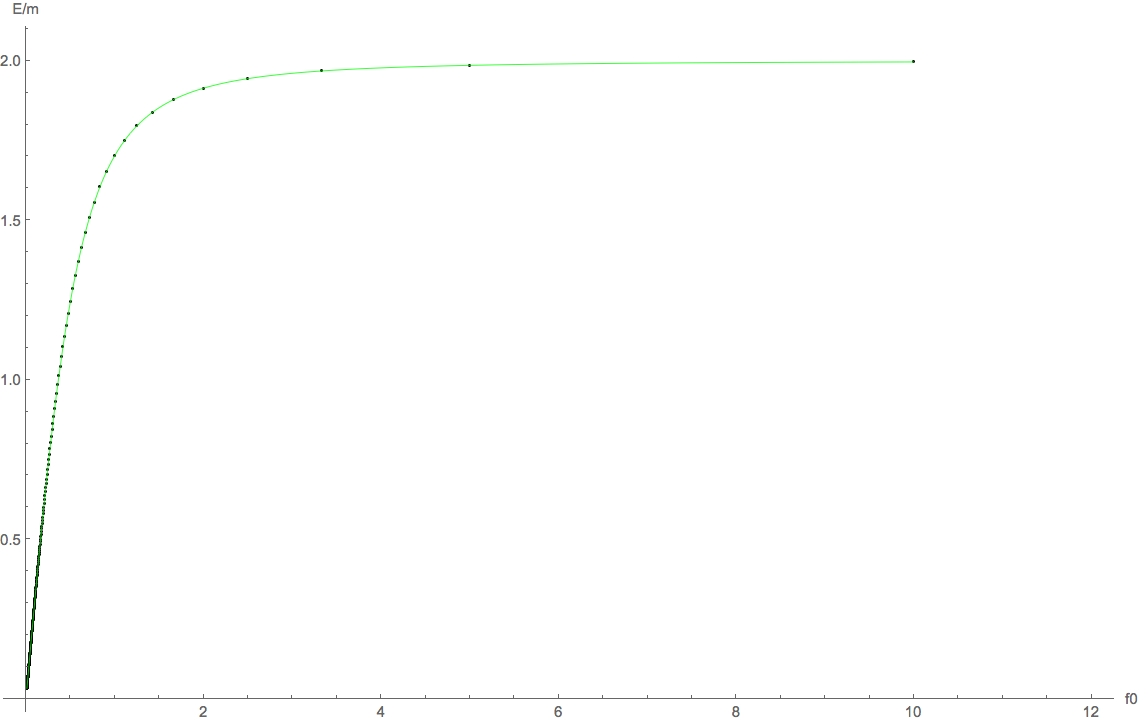}}
\quad
\subfloat[]{
\includegraphics[height=5cm,width=7cm]{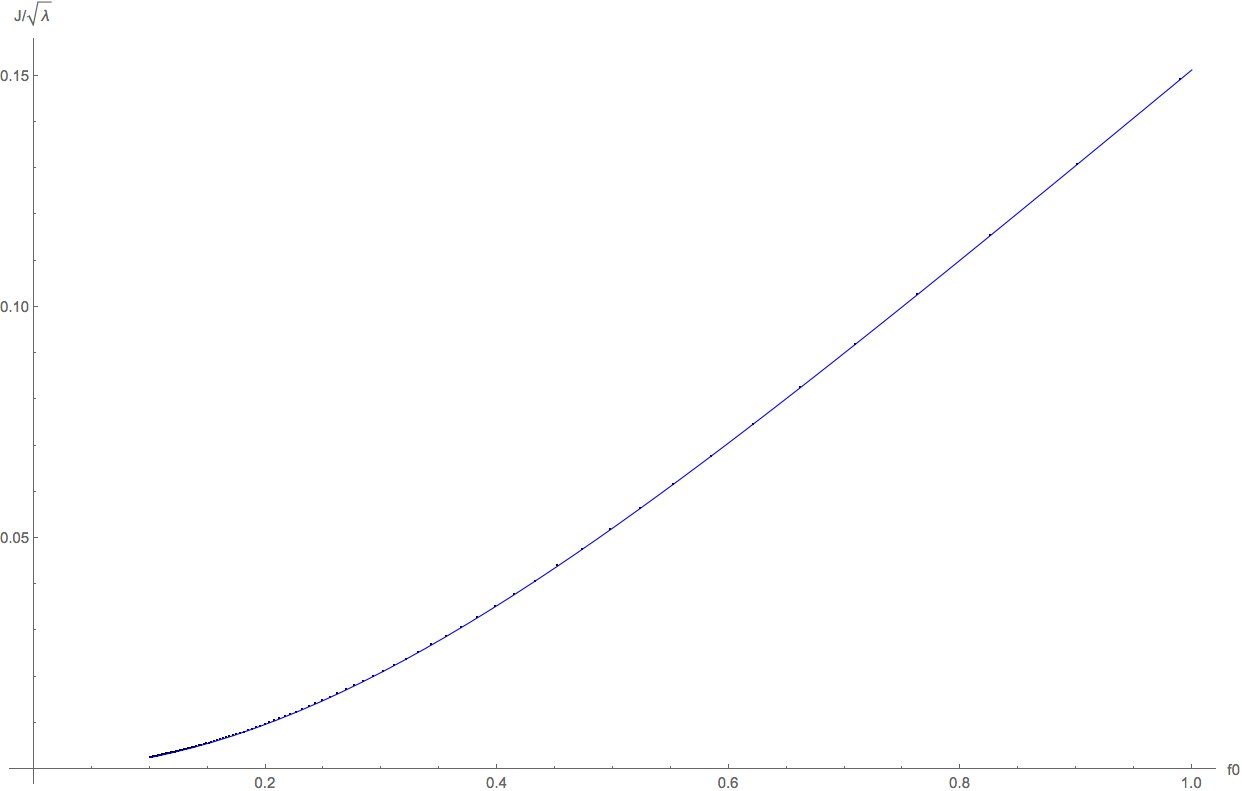}}}
\caption{\small\sffamily (a) The energy as function of the parameter $\tz_0$ for the meson model. The asymptotic limits are plotted in blue for $\tz_0<<1$ and red for $\tz_0>>1$. The dotted curve corresponds to the numerical integration of the profile of the string and numerical evaluation of the integral (\ref{ES}).  (b) The angular momentum as function of the parameter $\tz_0$ for the meson model. The asymptotic limits are plotted in blue for $\tz_0<<1$ and red for $\tz_0>>1$. The dotted curve correspond to the numerical integration of the profile of the string and numerical evaluation of the integral (\ref{JS}). (c) In green, the energy as function of the parameter $f_0$, $E/m_q=2\cos(\theta(f_0))$. The dotted curve correspond to the numerical integration of the profile of the string and numerical evaluation of the integral (\ref{Energy}) as function of $f_0$ after using our dictionary (\ref{dict}). (d) The angular momenta as function of $f_0$. The blue curve is the the plot of the relation (\ref{R2}). The dotted curve is an numerical integration of the profile of the string and the numerical evaluation of the integral (\ref{Spin}) in terms of $\tz_0$. Using our dictionary (\ref{dict}) the agreement is evident.}
\end{figure}

The other reason comes from the fact that two very different constructions, the rotating string on the one hand and the Wilson loop with a cusp on the other, are related using a very simple dictionary. From the side of the Wilson loop it is not clear where the complete profile of the string is encoded. From the side of the meson model it is not clear how the knowledge  of the cusp of the Wilson loop and the cusp anomalous dimension determines the  basic observables of the gauge theory to any order in the relevant parameters. That means that we are solving the differential equation for the string profile extending the asymptotic analysis that is reported in the current literature.

As we noted, the full lines (in colour) in Fig.~5 (c) and (d), are the plots of the numerical evaluation of the duality relation using information provided by the Wilson loop approach. So we do not need to know the complete profile of the string in the meson model to obtain the observables $E$ and $J$. What we need is only one point of the complete profile namely $\tilde{z_0}$ the maximum penetration of the string in the bulk that is also  related  to the position of the D7 brane $\tilde{z}_{D7}$ through (\ref{zd7z0pade}).

We also observe that the plot of the energy vs. angular momentum ($E-J$ graph reported previously in \cite{Caron-Huot-Henn} using numerical methods, see Fig. 6) does not change by the simple argument that the dictionary (\ref{dict}) is just  a change in the parametrization of $E$ and $J$. We can use as a parameter $f_0$ and $E(f_0), J(f_0)$ to obtain a point in this plot or we can use ${\tilde z_0}$ as a parameter, $E({\tilde z_0}), J({\tilde z_0})$ to obtain the complete plot and of course the graphs will be exactly the same. So we can reproduce the previous  numerical analysis with our simple dictionary. As a consequence of this fact we can see that from the side of  the meson model we need to integrate a complete profile of the string that is a solution of the NG equation of motion to obtain one point in this graph. Now we need only to know the maximum penetration of the profile (and not the complete profile!) to construct the entire plot $E-J$.

\section{Final comments}

We have proposed a new recipe for computing $E$ and $J$ for the meson spectrum using information from $\Gamma_{cusp}(\theta)$ and the duality (\ref{duality}) in the context of strong coupling, for the ${\cal N} = 4$ which have large spin (semiclassical limit), and whose gravity duals are semiclassical strings rotating in the AdS$_5\times$ S$_5$ space-time. Specifically, we constructed an explicit dictionary that relate the meson model observables with the observables of the Wilson loop with a cusp. We have a prediction at {\em any} order in $\tilde z_0$ for the observables of the meson model. Implicitly we have solved a differential equation (the profile of the string in the meson model) in terms of closed functions $\theta(f_0)$ and $\Gamma_{cusp}(\theta)$ that depend only on the point  $\tilde z_0$.

We have covered the case when the Wilson loop is in the  fundamental representation, a ``quark" with mass $m_q$. Could be also worth explore the generalisation to the case of D3-branes \cite{Guijosa, Diego-Trancanelli}  (symmetric representation) and D5-brane (antisymmetric representation). We plan to investigate  this issue in a future work.

A holographic computation of the entanglement entropy in conformal field theories has been proposed via the AdS/CFT correspondence \cite{entangled}. In \cite{entangled1} the authors examine the strong subadditivity constraint via direct calculations. As an example they work out the Wilson loop with a cusp and confirm strong subadditivity. The calculation is analogous to the one reviewed here (see Section 3). It is interesting to note that this implies also via the dictionary (\ref{dict}) that the quark-antiquark pair is entangled. In fact, in \cite{entangled1} the authors find that the entanglement entropy can be obtained as a direct application of the procedure to find $\Gamma_{cusp}$.
As we know, the total area is
$$A /R^2= 2L/\epsilon - \Gamma_{cusp}(\theta) \log L/\epsilon + {\hbox {(finite terms)}}. $$
where $R$ is the AdS radius.

In \cite{entangled}, it is claimed that the entanglement entropy can be computed as follows
$$ S_A = \frac{A(\Sigma)}{ 
4G_N^{(d+2)}}$$
where $A(\Sigma)$ denotes the area of the surface $\Sigma$, and $G^{(d+2)}_N$ is the Newton constant in 
$d + 2$ dimensional AdS space. The $d$ dimensional surface $\Sigma$ is determined in such way that is the minimal area surface whose boundary coincides with the boundary of the submanifold A. Thus the entanglement entropy can be  computed as (up to constant terms)
$$ S_A= \frac{R^2}{4 G_N^4}  \Big(2L/\epsilon-\Gamma_{cusp} \log L/\epsilon\Big) $$
So our guess is that entanglement entropy of the quark anti-quark pair can be written in terms of the spin $J(\tilde z_0)$ of the meson in the gauge theory. In this way the entanglement entropy depends only on the parameter $\tz_0$ of the dual string.

Recently in \cite{entangled2} the authors found that the entanglement entropy of three-dimensional conformal field theories contains a universal contribution coming from corners in the entangling surface. This calculation is also very similar to that of $\Gamma_{cusp}$. It could be interesting to study these results from the point of view our dictionary to obtain (universal?)  physical information encoded in the meson model.

A fundamental understanding of the proposed duality is still lacking. We have no rigorous justification for the interpretation or a convincing geometrical argument for how it works out so well. To answer such question we could need to relate the two worldsheets and not only the two parameters $\tilde z_0$ and $f_0$.
 Also, we do not know whether it would make sense, for more general strings, undergoing some complicated motion (e.g. an infinite string with one endpoint attached to an accelerated heavy quark), and what should be the physical interpretation in that case.
 
 We hope that such questions will trigger further investigations, leading to conceptual clarifications and to new results.
   
 For completeness, let us finally mention that corresponding to open string configurations with large angular momentum have been considered within the AdS/CFT correspondence, but for a different space-time geometry, 
and/or different dynamical situations. It is natural to wonder if these cases can be approached from the same point of view of this article.

\section{Acknowledgements}

We are grateful to Alberto G\"uijosa, for useful discussions and suggestions on the manuscript. The authors was  partially supported by Mexico National Council of Science and Technology (CONACyT) grant 104649 and DGAPA-UNAM grant IN110312.

\begin{appendices}

\section{Asymptotic limits for the cusp angle integral (\ref{thetaintK})}

We can estimate the value of $\theta(f_0)$ of the integral (\ref{thetaintK}) in two relevant limits. These asymptotic behaviour is important because we can make contact with the corresponding limits of the meson energy and angular momenta through our dictionary (\ref{dict}).   The limit $f_0 \rightarrow 0$ of the integral
\be\nonumber
 \theta(f_0) = 2 K \int\limits_0^\infty \dfrac{d\eta}{(\eta^2 + f_0^2)\sqrt{(\eta^2+f_0^2+1)(\eta^2+2f_0^2+1)}}, 
\ee
can be estimated using an appropriate Taylor expansion observing that any divergence is an artefact of the approximation. As we already has been observed the integral is a smooth function of the parameter $f_0$, in fact is a monotonically decreasing function of $f_0$.
In the  limit $f_0 \ll 1$, after exploring the general behaviour of the integrand around $f_0\to 0$ we can estimate
\be
\theta(f_0 ) = -\frac{36305 \pi  f_0^9}{16384}+\frac{359 \pi  f_0^7}{256}-\frac{61 \pi  f_0^5}{64}+\frac{3 \pi  f_0^3}{4}-  \pi  f_0 +\pi \cdots.
\label{thetaserie}
\ee
To show explicitly that this estimation works very well see Fig. 6 (green curve).

From the other hand  in the limit $f_0 \gg 1$ the integral (\ref{thetaintK}) behaves like
 \be
 \theta(f_0) = \frac{2 \pi^{1/2} \Gamma\left( \frac{3}{4} \right) }{\Gamma\left( \frac{1}{4} \right) } \frac{1}{f_0}-\frac{0.08}{f_0^3} + \frac{0.005}{f_0^5} + \calo(\frac{1}{f_0^7}),
 \label{thetaserielarge}
 \ee
The estimates (\ref{thetaserie}, \ref{thetaserielarge}) covers practically all the intermediate values of the integral as can be appreciated  in Fig. 6.

\begin{figure}[h]
\begin{center}
\includegraphics[height=9cm,width=14cm]{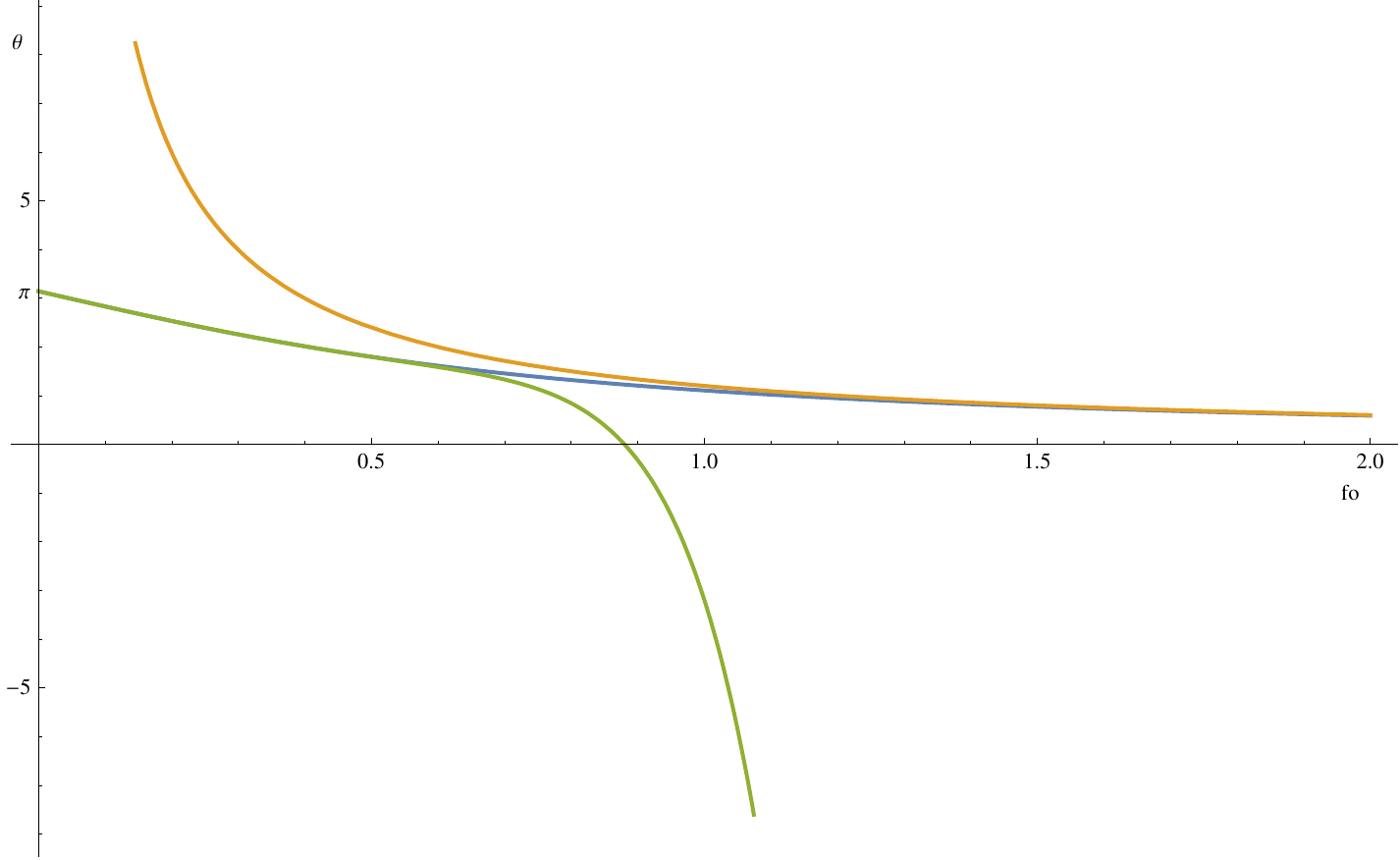}
\caption{ \small\sffamily The numerical result of the integral (\ref{thetaintK}) is in blue. The other graphs are the estimation of the asymptotic  behaviour of the integral given in (\ref{thetaserie},\ref{thetaserielarge}).}
\end{center}
\end{figure}

\end{appendices}

\newpage

\providecommand{\href}[2]{#2}\begingroup\raggedright\endgroup

\end{document}